\documentclass[prl,preprint,superscriptaddress,longbibliography]{revtex4-1}
\usepackage{graphicx}
\usepackage{hyperref}
\usepackage{soul}
\usepackage{amsmath}
\usepackage[usenames,dvipsnames]{color}

\begin{document}

\title{Coupled Yu-Shiba-Rusinov states in molecular dimers on NbSe$_2$}

\author{Shawulienu Kezilebieke}
\affiliation{Department of Applied Physics, Aalto University School of Science, P.O.Box 15100, 00076 Aalto, Finland}

\author{Marc Dvorak}
\affiliation{Centre of Excellence in Computational Nanoscience (COMP) and Department of Applied Physics, Aalto University, P.O.Box 11100, 00076 Aalto, Finland}

\author{Teemu Ojanen}
\email{Email: teemuo@boojum.hut.fi}
\affiliation{Department of Applied Physics, Aalto University School of Science, P.O.Box 15100, 00076 Aalto, Finland}

\author{Peter Liljeroth}
\email{Email: peter.liljeroth@aalto.fi}
\affiliation{Department of Applied Physics, Aalto University School of Science, P.O.Box 15100, 00076 Aalto, Finland}

\begin{abstract}
Magnetic impurities have a dramatic effect on superconductivity by breaking the time-reversal symmetry and inducing so-called Yu-Shiba-Rusinov (YSR) low energy bound states within the superconducting gap. The spatial extent of YSR states is greatly enhanced in 2D systems, which should facilitate the formation of coupled states. Here, we observe YSR states on single cobalt phthalocyanine (CoPC) molecules on a 2D superconductor NbSe$_2$ using low-temperature scanning tunneling microscopy (STM) and spectroscopy (STS). We use STM lateral manipulation to create controlled CoPc dimers and demonstrate the formation of coupled YSR states. The experimental results are corroborated by theoretical analysis of the coupled states in lattice and continuum models. Our work forms an important step towards the realization of exotic topological states in designer magnetic lattices. 
\end{abstract}

\maketitle
\newpage

Magnetic impurities have a dramatic effect on superconductivity by breaking the time-reversal symmetry and inducing so-called Yu-Shiba-Rusinov (YSR) low energy bound states within the superconducting gap \cite{Shiba1968,LUH1965,Rusinov1969,Balatsky2006,Yazdani1997}. These states can be detected in real space by scanning tunneling microscopy (STM) and their energy spectrum has been studied in great detail by scanning tunneling spectroscopy (STS) on a variety of traditional s-wave superconductors \cite{Yazdani1997,Ji2008,Franke2011,Ruby2015,Hatter2015,Ruby2016,Wiebe2017}. In addition to individual impurities, self-assembled dimers at atomic separations \cite{Ji2008} and atomic wires of magnetic atoms have been investigated \cite{Nadj-Perge2014,Ruby2015a,Franke2017}.

Ferromagnetic coupling between the two impurities results in  YSR states that hybridize and split into bonding and anti-bonding states \cite{Flatt2000, Morr2006, Morr2003, Ji2008}. Atomic chains with a suitable spin-texture have been suggested to support 1D topological superconductivity \cite{Nadj-Perge2013} and Majorana modes at each end \cite{Kitaev2001}, which have been recently realized in iron chains on Pb(110) \cite{Nadj-Perge2014,Ruby2015a}. In general, controlled coupling of YSR states should enable realizing designer quantum materials with novel topological phases \cite{Rontynen2015}. However, the experimentally observed YSR states on s-wave superconductors (Pb and Nb) are localized within a few atomic distances from the impurity centre and their energies vary strongly depending on the adsorption site of the impurity. This greatly hampers forming assemblies of controllably coupled YSR states.

Recently, M\'{e}nard et al. demonstrated that the spatial extent of the YSR states depends on the dimensionality of the superconductor and is greatly enhanced in 2D systems, where the impurity bound state can extend over several nanometres away from the impurity \cite{Menard2015}. This was demonstrated on iron impurities embedded into a niobium diselenide (NbSe$_2$) substrate, which is layered material with a superconducting transition at 7.2 K. Due to the weak van der Waals interaction between adjacent layers, it behaves essentially as a 2D system. However, coupling between YSR states on 2D superconductors has not yet been demonstrated. Here, we observe YSR states on single cobalt phthalocyanine (CoPC) molecules on a 2D superconductor NbSe$_2$ using low-temperature scanning tunneling microscopy (STM) and spectroscopy (STS). We use STM lateral manipulation to create controlled CoPc dimers and demonstrate the formation of coupled YSR states. The experimental results are corroborated by theoretical analysis of the coupled states in lattice and continuum models. Our work forms an important step towards the realization of exotic topological states in designer magnetic lattices \cite{Rontynen2015,Menard2017}.

\begin{figure}[!t]
\centering
\includegraphics [width=0.75\textwidth] {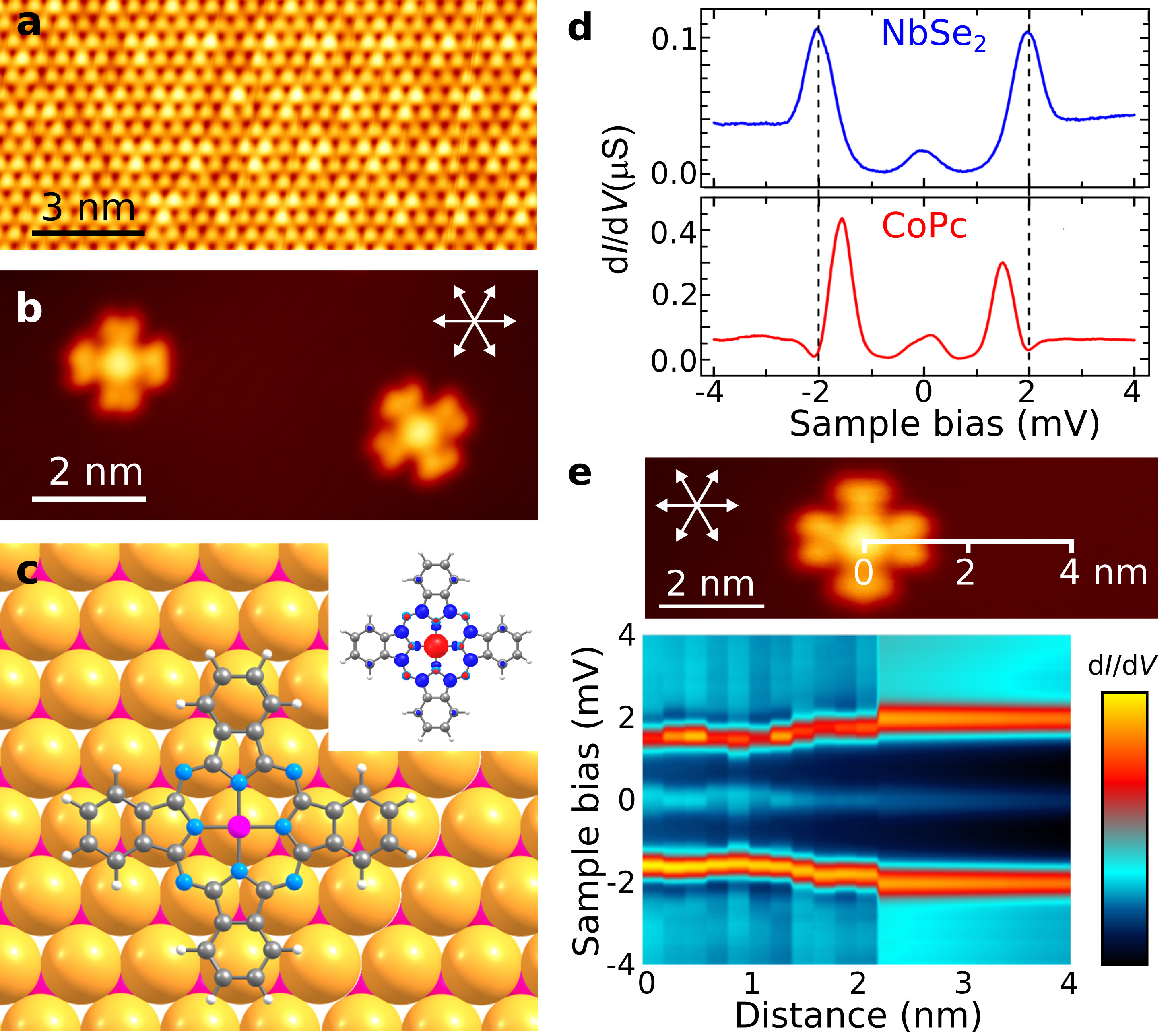}
\caption{CoPc molecules deposited on NbSe$_2$ surface. (a) Atomically resolved STM image of the NbSe$_2$ surface ($V=20 $ mV, $I=1$ nA). (b) STM image of two CoPc molecules ($V=0.6$ V, $I=5$ pA). The arrows indicate the principal directions of the underlying NbSe$_2$ substrate. (c) DFT results on the CoPc adsorption on NbSe$_2$. Inset shows the calculated spin density on the CoPc molecule. (d) Measured d$I$/d$V$ spectra on the bare NbSe$_2$ substrate (top panel, feedback setpoint: $V=100$  mV, $I=50$  pA, $z_\mathrm{offset}=100$ pm) and on an isolated CoPc molecule (bottom panel, feedback setpoint: $V=100$  mV, $I=50$  pA, $z_\mathrm{offset}=50$ pm) with a superconducting STM tip. (e) d$I$/d$V$ spectra acquired at different distances from the centre of a CoPc molecule showing the evolution of the YSR resonances (feedback setpoint: $V=100$  mV, $I=50$  pA, $z_\mathrm{offset}=50$ pm). Colour scale between $0-1$ $\mu$S. In the upper panel, the arrows indicate the principal directions of the underlying NbSe$_2$ substrate.}
\label{fig1}
\end{figure}

Figure \ref{fig1}a shows an atomic resolution STM image of a NbSe$_2$ surface at 4.2 K (See Supporting Information for details on the experiments). It reveals the hexagonal arrangement of the outermost Se atoms, and the well-known $3\times3$ charge-density wave (CDW) superstructure \cite{Pan1998,Menard2015}. Figure \ref{fig1}b displays a topographic STM image of two isolated cobalt phthalocyanine (CoPc) molecules on NbSe$_2$, where the principal directions of the substrate have been marked with arrows. We have determined the adsorption site of CoPc from atomically resolved images (Supplementary Information), which is relevant as the molecular properties (e.g. spin state and more importantly, the magnetic coupling to the superconductor) might depend on this. Our STM measurements indicate that independent of the molecular orientation, the central Co atom always sits on top of an underlying Se atom. This experimentally observed adsorption configuration is confirmed by density-functional theory (DFT) calculations, with Fig. \ref{fig1}c showing the most stable adsorption structure.  

d$I$/d$V$ spectroscopy was performed with NbSe$_2$-coated SC tip to increase the energy resolution beyond the thermal limit \cite{Pan1998,Ji2008,Franke2011}. Figure \ref{fig1}d (upper panel) shows a typical d$I$/d$V$ spectrum acquired on bare NbSe$_2$ substrate. It exhibits a gap that is twice the size of the SC gap of NbSe$_2$, with sharp coherence peaks at $\pm 2\Delta \sim \pm 2$ meV \cite{Noat2015, Rodrigo2004}. The small peak close to $V=0$ is due to the tunneling of thermally excited quasiparticles. d$I$/d$V$ curve taken on a CoPC molecule (Fig. \ref{fig1}d, lower panel) shows that the coherence peaks at $\pm 2\Delta$ are replaced by pronounced peaks of asymmetric heights within the SC gap and two replicas of those peaks close to $V=0$, similarly to an earlier report on MnPC on Pb surface \cite{Franke2011}. These asymmetric peaks in the SC gap arise from the interaction between an isolated spin on CoPC and Cooper pairs in NbSe$_2$, i.e. the formation of YSR states. Despite CoPc adsorption being driven by the weak van der Waals forces, the magnetic coupling is still sufficiently strong to result in the formation of YSR states. DFT calculations predict that CoPC on NbSe$_2$ has the same spin as in the gas phase, $S=1/2$. The inset of Fig. \ref{fig1}c shows the spin density of CoPC on NbSe$_2$ surface, which is derived from the cobalt $d_{z^2}$ orbital, as expected.  

Because of the 2D nature of NbSe$_2$, the YSR states have been shown to have a large spatial extent and to decay more slowly than on a 3D superconducting substrate\cite{Menard2015}. We have probed the spatial extent of the YSR states in our system by recording spectra along a line over a single CoPc as shown in Fig. \ref{fig1}e (zero corresponds to the centre of the molecule). It is seen from the figure that the YSR states persist $> 2$ nm from the centre of the molecule. In addition to the very slight energy variations of the YSR resonance over the molecule (distances $< 1$ nm), the smooth energy variation of the resonances at larger distances is likely to result from tip-molecule interactions (see Supplementary Information for more details). The YSR wavefunctions on NbSe$_2$ are expected to have six-fold symmetry with larger spatial extent along the principal crystal directions of the surface\cite{Menard2015}. We have tested this effect by recording another set of line spectra at a $45^\circ$ angle with respect to the data shown in Fig. 1e, and carried out grid spectroscopy experiments (Supplementary Information). These experiments suggest that YSR wavefunctions are anisotropic also in our system.

\begin{figure}[!h]
\centering
\includegraphics [width=0.95\textwidth] {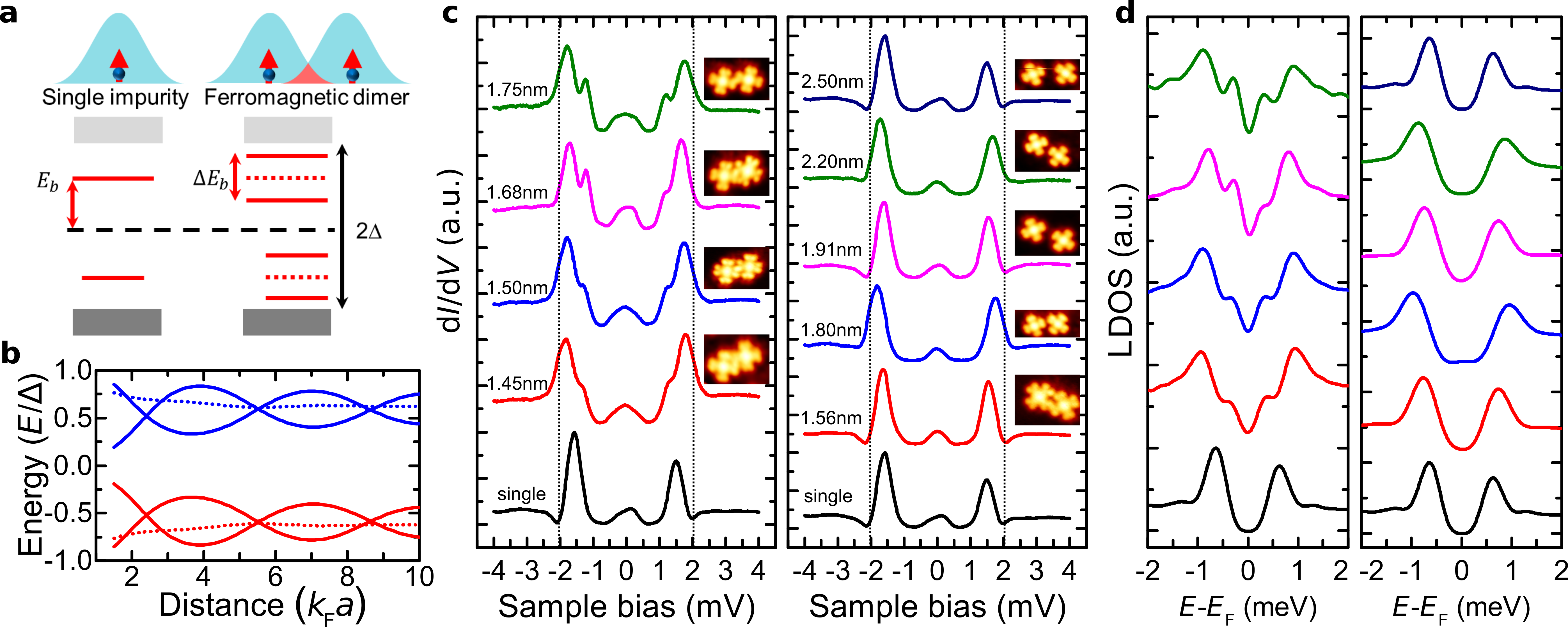}
\caption{Formation of coupled YSR states on CoPc dimers. (a) Schematic of the formation of coupled YSR states. (b) Calculated evolution of the YSR state energies as a function of the impurity dimer separation for a ferromagnetic (solid lines) and antiferromagnetic (dotted lines) dimer. (c) Set of d$I$/d$V$ spectra showing split (left panel) and non-split YSR states (right panel). A spectrum measured on an isolated CoPc (black line) is shown in both panels for comparison (feedback setpoint: $V=100$  mV, $I=50$  pA, $z_\mathrm{offset}=100$ pm). The dotted line shows the position of the SC gap edge at $\pm2\Delta/e$. (d) The LDOS extracted by numerical convolution from the experimental d$I$/d$V$ curves.}
\label{fig2}
\end{figure}

The coupling of subgap states is illustrated in Fig. \ref{fig2}a, which shows schematically how two nearby YSR states hybridize and form bonding and antibonding combinations. In this case, phenomenological theory predicts a splitting that depends on the dimensionless coupling strength $\alpha$ (i.e the energy of an individual YSR state), the relative orientations of the two (classical) spins and the separation between the spins \cite{Flatt2000,Balatsky2006,Meng2015,Rontynen2015} (Supplementary Information). $\alpha$ can be determined from experiments on a single impurity and we estimate a value of $\alpha\approx0.5$. The splitting oscillates with a period determined by $k_F$ (See Fig. \ref{fig2}b) and obtains its maximum value for a ferromagnetic alignment of the spins. Coupling of antiparallel spins (dotted line in Fig. \ref{fig2}b) does not result in energy splitting but the degenerate energy level is slightly shifted from the individual YSR energy.

Using STM lateral manipulation \cite{Eigler1990}, we successfully constructed molecular dimers with different separations (see Supplementary Information for details). In all cases, we have manipulated one molecule of the dimer and have recorded the spectrum on the one which has not been moved. In this way, we have made sure that the target molecule is always at the same adsorption side. Figure \ref{fig2}c shows point spectra on the molecular dimers with different separations (STM images of the dimers are shown in the insets). Unlike in the case of a single CoPc molecule where we observed only one pair of YSR resonances in the SC gap, now, the main YSR peaks are split to two peaks (left panel). In order to remove the influence of the superconducting density of states of the tip, numerical deconvolution was performed to directly extract the local density of states (LDOS) of the YSR states due to the CoPc molecules. Figure \ref{fig2}d shows the measured LDOS, where the split YRS states (left panel) can be compared to the single YSR peaks at $\pm$0.63 meV. It is evident that the energy positions of split peaks change depending on distance between the dimers. 

We observe a maximum splitting of $\sim0.5$ meV, which is very similar to the predictions from a simple continuum model for experimentally relevant distances (Fig. \ref{fig2}b). Magnetic coupling between the spins (through e.g. RKKY interaction or superexchange) is expected to be weak compared to $kT$ at $T=4.2$K. Therefore the spins in a dimer are randomly oriented (thermal average). Since majority of random orientations give rise to energy splittings comparable to the maximum splitting, we would observe experimentally essentially the same splitting as the maximum value (see Supplementary Information for details).

Other possible reasons for splitting of the YSR states include different angular momentum scattering contributions, individual $d$ orbitals acting as separate scattering potentials or low-energy excitations due to magnetic anisotropy or vibrations \cite{Kunz1980, Flatt1997, zitko2011, Golez2012,Ruby2015,Hatter2015,Ruby2016}. All of the above scenarios are predicted on a single magnetic impurity, where we always observe only a single pair of YSR states. This leaves the magnetic coupling between the impurities through the SC medium and the formation of coupled YSR states as the natural explanation of the observed spectra. 

The right panel of Figure \ref{fig2}c shows a different set of d$I$/d$V$ spectra on the molecular dimers. While not identical, the intra-dimer distances are over a similar range as in left panel of Fig. \ref{fig2}c.  Surprisingly, small changes in the distance result in drastic changes in the spectra and we observe alternating single and split YSR state behaviour. Once the impurity separations exceed $\sim 2.5$ nm, we only observe  response consistent with single impurity YSR states. As indicated above, we do not expect the spins to be strictly antiparallel, and the expected distance dependence (period of $k_F$) is not consistent with these rapid changes between split and non-split dimer states.

\begin{figure}
\centering
\includegraphics [width=0.95\textwidth] {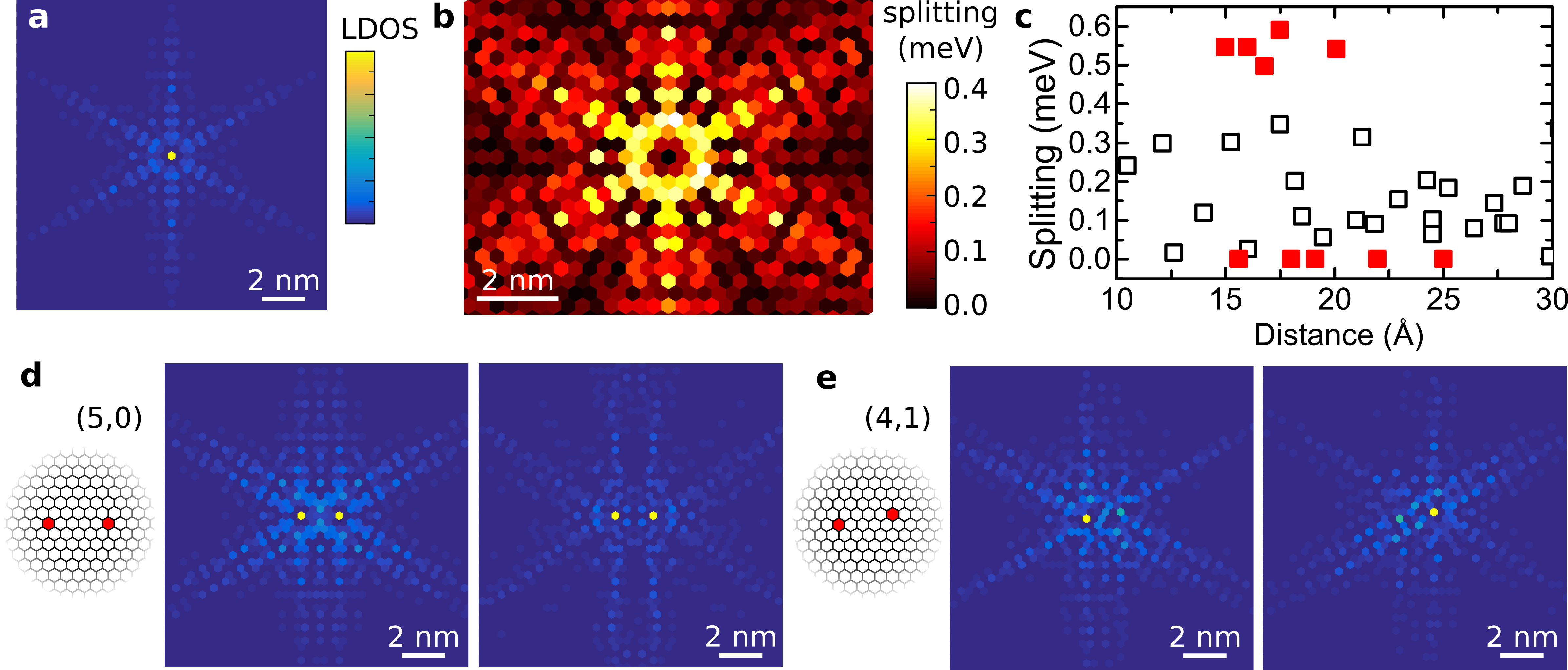}
\caption{Atomic scale details of the coupling between two magnetic impurities. (a) Calculated LDOS of single impurity in the next-nearest neighbour tight-binding model. (b) Calculated splitting as a function of the impurity position (the other impurity is at (0,0)). (c) Calculated (open symbols) and measured (closed symbols) splitting of the YSR states as a function of the distance between the impurities. (d,e) Calculated bonding (left panels) and antibonding (right panels) state LDOS for a strongly coupled (5,0) (panel d) and weakly coupled (4,1) (panel e) dimer.}
\label{fig3}
\end{figure}

In order to understand this behaviour, we have to go beyond the continuum description of the SC substrate. The Fermi surface of the NbSe$_2$ is anisotropic \cite{Rossnagel2001}, which gives rise to star-shaped structures in the LDOS around a magnetic impurities and vortices \cite{Menard2015}. In addition, the YSR wavefunctions have atomic scale oscillations arising from the Bloch part of the wavefunction. As the coupling of the YSR states is determined by the wavefunction overlap, these oscillations are a potential source for the atomic scale variations in the observed YSR splitting (see below). We model these effects using a next-nearest neighbour tight-binding model \cite{Flatt2000} (see Supplementary Information for details). Calculated LDOS of a single impurity is shown in Fig. \ref{fig3}a. The wavefunction strongly reflects the crystal symmetry and it is easy to see that the coupling might strongly depend on the relative orientation of the dimer. We have calculated coupled dimers for different positions of the magnetic impurities and the splitting of the YSR states is shown in Fig. \ref{fig3}b. In addition to a very clear six-fold symmetry stemming from the crystal lattice, there are strong atomic scale variations. This variations result from strong changes in the wavefunction overlap (due to the Bloch part of the wavefunction), when one of the impurities is moved by a single lattice site. This is highlighted in Fig. \ref{fig3}c, which shows the YSR splitting as a function of the dimer separation over the experimentally relevant range. It is seen to oscillate wildly, in agreement with the experimental data (shown with solid red squares).

In Fig. \ref{fig3}d,e we have illustrated the bonding and antibonding wavefunctions of two dimers, where one of the impurities has been moved by a single lattice site from (5,0) to (4,1) (the other impurity is at (0,0)). The calculated YSR splitting in these two dimers is 0.35 meV (5,0) and 0.02 meV (4,1). The latter is far below our experimental energy resolution and would not result in an observable splitting of the YSR resonances. The calculated wavefunctions reflect this: while they are delocalized on both impurities in the strongly coupled dimer (Fig. \ref{fig3}d, analogous to H$_2$ molecule), they are mostly localized on a single impurity in the weakly coupled dimer (Fig. \ref{fig3}e). This highlights that the atomic scale details are important for a detailed understanding of the coupled YSR states.

In conclusion, we have demonstrated the formation and coupling of YSR states on CoPc molecules on NbSe$_2$. Using STM lateral manipulation, we have constructed well-defined CoPc dimers and observed coupled YSR states. Experimentally, we find strong variations of the coupling strength depending on the detailed geometry of the CoPc dimer, which can be understood based on the details of the wavefunction overlap of the two impurity states. The demonstration of coupled YSR states is promising for realization of novel topological states predicted in YSR lattices.

\section*{Acknowledgements}
This research made use of the Aalto Nanomicroscopy Center (Aalto NMC) facilities and was supported by the European Research Council (ERC-2011-StG No. 278698 “PRECISE-NANO”), the Academy of Finland through its Centres of Excellence Program (projects no. 284594 and 284621) and the Academy Research Fellow (T.O., No. 256818) and Postdoctoral Researcher (S.K., No. 309975) programs, and the Aalto University Centre for Quantum Engineering (Aalto CQE).

\bibliography{Shiba}

\newpage

\renewcommand\thefigure{S\arabic{figure}} 
\setcounter{figure}{0}
\section*{Supplementary Information}

\section*{Methods}
\textbf{Sample preparation.} Sample preparation and subsequent STM experiments were carried out in an ultrahigh vacuum system with a base pressure of $\sim$10$^{-10}$ mbar. The $2H$--NbSe$_2$ single crystal (HQ Graphene, the Netherlands) was cleaved in situ by attaching a tape to the crystal surface and pulling the tape in vacuum in the load-lock chamber using the sample manipulator. CoPC molecules (Sigma-Aldrich) was deposited from an effusion cell held at 390$^{\circ}$C onto a freshly cleaved NbSe$_2$ at room temperature. 

\textbf{STM measurements.} After the CoPc deposition, the sample was inserted into the low-temperature STM (Unisoku USM-1300) and all subsequent experiments were performed at $T=4.2$ K. STM images were taken in the constant current mode. d$I$/d$V$ spectra were recorded by standard lock-in detection while sweeping the sample bias in an open feedback loop configuration, with a peak-to-peak bias modulation of $50-100$ $\mu$V at a frequency of 709 Hz. The procedure for acquiring a spectrum was as follows: the tip was moved over the molecule at the imaging parameters (e.g. $V = 0.6$ V and $I = 5$ pA), the tip-sample distance was reduced by changing the setpoint to e.g. 200 pA at 100 mV. Finally, after disconnecting the feedback at the beginning the d$I$/d$V$ spectrum, the tip-sample distance was decreased by a further $50 - 100$ pm ($z_\mathrm{offset}$) to increase the signal to noise ratio. The detailed numbers are given in the figure captions.

The NbSe$_2$ tip was prepared by indenting the tip into the NbSe$_2$ surface by a few nanometers while applying a voltage of 10 V. Manipulation of the CoPc was carried out by placing the tip above the centre of the molecule with a bias voltage of 0.1 V and the current was increased to 1 nA with the feedback engaged. The tip was then dragged towards the desired location.

\textbf{DFT calculations.} Density functional theory calculations were performed with the FHI-AIMS computational package \cite{Blum2009,Havu2009} and the PBE generalized gradient approximation for the exchange-correlation functional \cite{Perdew1996}. We used the standard "light" numerical settings and basis sets of numeric atomic-centered orbitals tested and recommended by FHI-AIMS. Periodic NbSe$_2$ supercells were sampled with a $2\times2$ k-point grid centred on the $\Gamma$ point. Van der Waals interactions were included by the post-SCF Tkatchenko-Scheffler correction \cite{Tkatchenko2009}. Before computing the electronic structure, all atomic forces were relaxed to $<0.01$ eV/{\AA}.

\section*{Adsorption geometry of CoPc}
\begin{figure}[!h]
	\centering
	\includegraphics [width=0.9\textwidth] {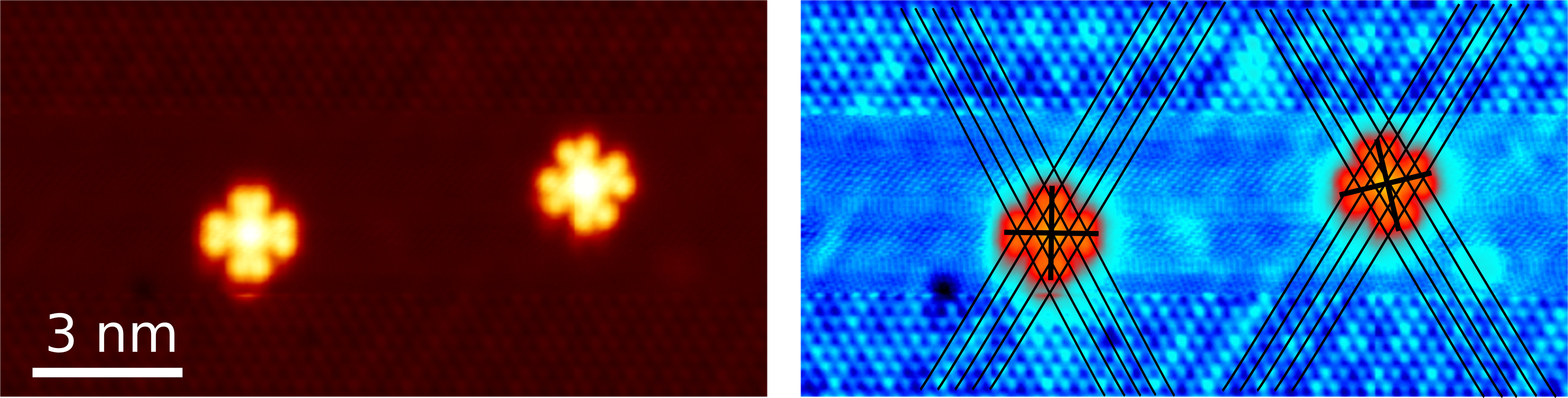}
	\caption{STM image of two CoPc molecules ($V=50$ mV, $I=1$ nA on NbSe$_2$ and $V=0.7$ V, $I=3$ pA on CoPc). The STM feedback current was reduced of the CoPc to avoid their accidental manipulation.}
	\label{FigSI2}
\end{figure}
The adsorption site of CoPc can be determined from atomically resolved STM images (Fig. \ref{FigSI2}). The STM feedback current was increased at the top and bottom parts of the image to allow resolving the NbSe$_2$ lattice. While scanning over the molecules, the setpoint current was reduced in order not to accidentally manipulate the molecules. Extrapolating the atomically resolved NbSe$_2$ lattice (Se is visible) allows estimating the CoPc adsorption site. Both molecules are adsorbed with the cobalt centre directly on top of a selenium atom in agreement with the DFT calculations.

\section*{Spectra over a single CoPc molecule}
\begin{figure}[!h]
	\centering
	\includegraphics [width=0.85\textwidth] {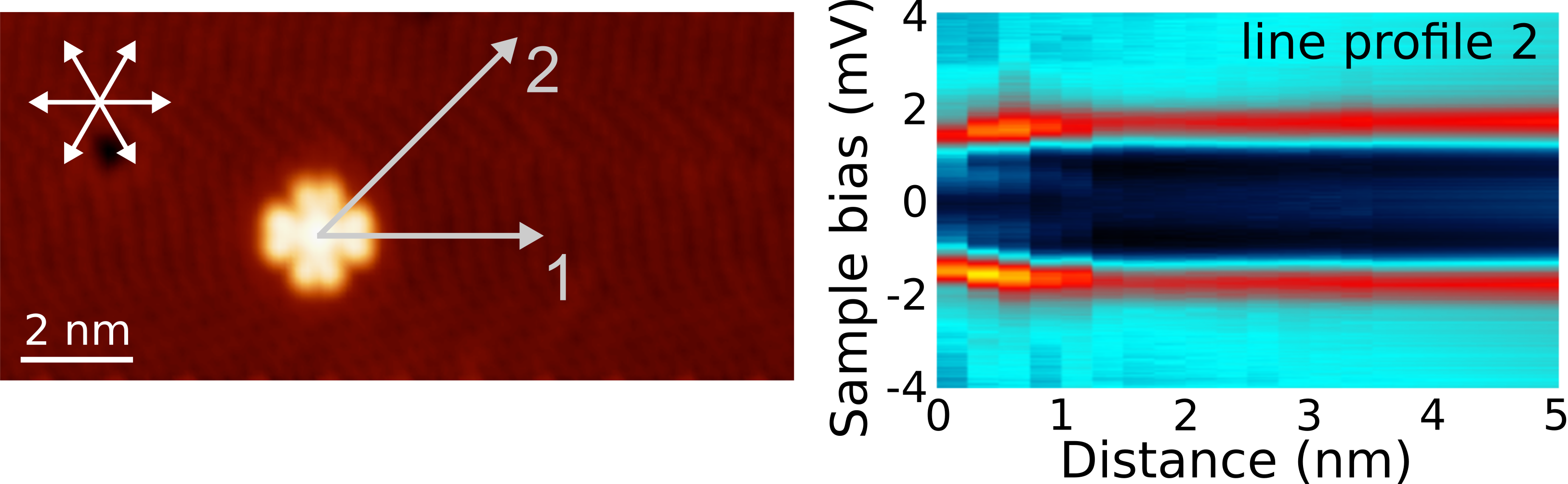}
	\caption{Spectra along a line over a CoPc molecule (profile 2). Spectra along line 1 are shown in Fig. 1. Feedback setpoint: $V=100$  mV, $I=200$  pA, $z_\mathrm{offset}=0$. The colour scale is between $0-0.03$ $\mu$S. }
	\label{FigSI2}
\end{figure}

\begin{figure}[!h]
	\centering
	\includegraphics [width=0.65\textwidth] {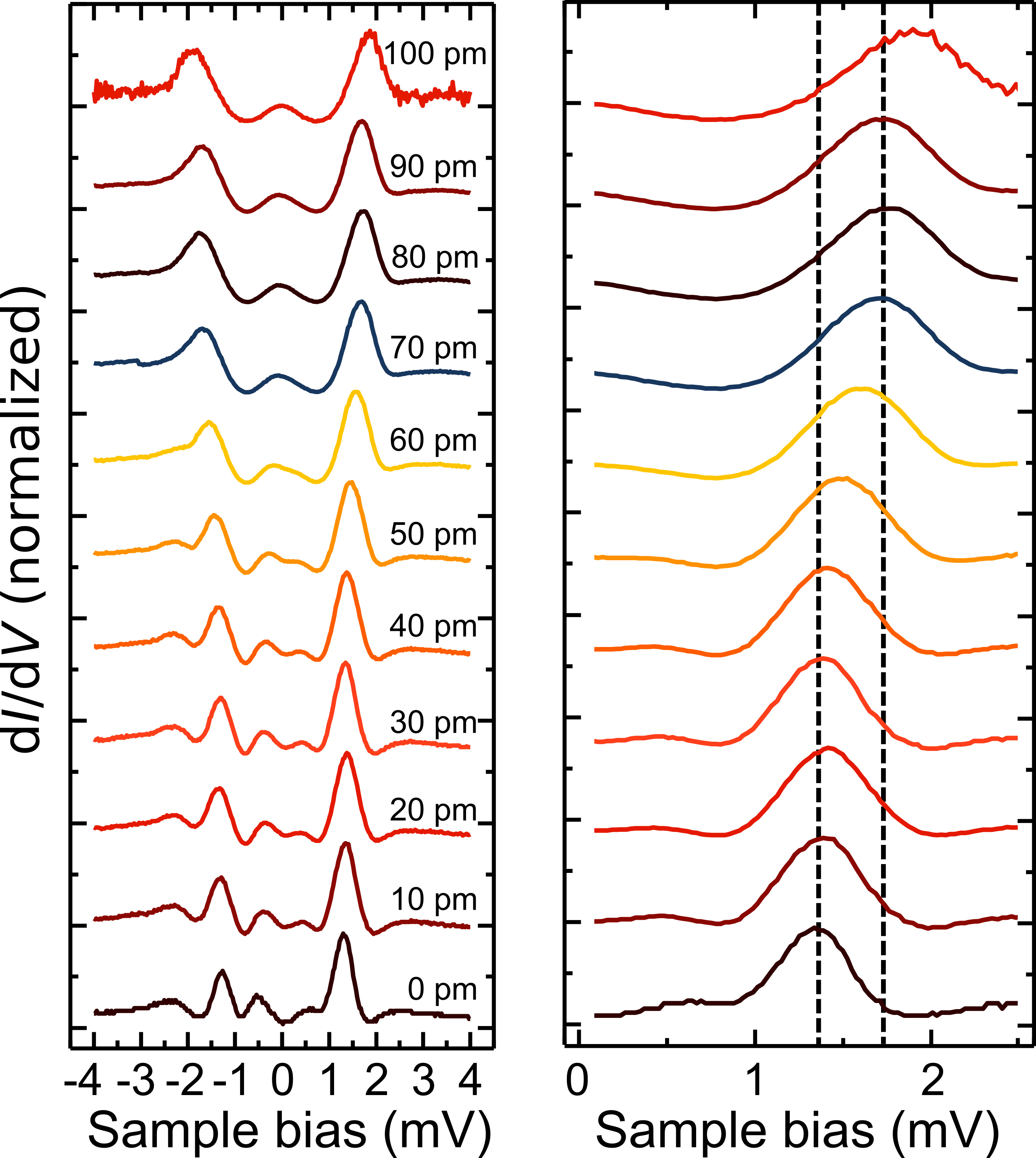}
	\caption{d$I$/d$V$ spectra (normalized) recorded over the middle of a CoPc molecule at different tip-molecule distances. $z_\mathrm{offset}=0$ pm corresponds to the set-point conditions at $V=20$  mV, $I=300$  pA. }
	\label{FigS_setpoint}
\end{figure}

We have carried out experiments to probe the nature of the individual YSR states. In addition to the spectra along line profile 1 shown in the main manuscript, we have additional data along profile 2 at a 45$^\circ$ angle w.r.t. profile 1 (Fig. \ref{FigSI2}). This data suggests that the YSR state has shorter decay along this direction. It can also been seen that the YSR resonances shift towards the gap edge, which is difficult to consolidate with the picture of the YSR states being eigenstates of the impurity-surface complex. This effect is also seen (to a lesser extent) on the profile 1 in Fig.~1e of the main text. To check if this effect could be caused by the interaction with the STM tip, we have measured spectra at different tip-sample distances in the middle of a CoPc molecule (Fig. \ref{FigS_setpoint}). We start all the experiments at a distance determined by the set-point conditions and approach the tip by a distance of $z_\mathrm{offset}$ before recording the spectrum.

d$I$/d$V$ spectra acquired at different tip-sample distances show a shift of the YSR resonance to higher bias starting at roughly $z_\mathrm{offset}= 30-40$ pm. At around $z_\mathrm{offset}= 80$ pm, the YSR resonance has merged with the superconducting coherence peak at the gap edge. The spectrum recorded at $z_\mathrm{offset}= 100$ pm has significantly higher noise and broader resonances compared to the other spectra, and the molecule becomes unstable under the tip at larger values of $z_\mathrm{offset}$. On the bare substrate, varying the tip-sample distance has no effect on the shape or position of the superconducting coherence peaks.

This measurement clearly indicates that there are considerable interactions between the tip and the CoPc molecules at reduced tip-sample distances. This interaction modifies the coupling with the underlying NbSe$_2$ substrate, as evidenced by the continuous shift of the YSR resonances as a function of the tip-sample distance. Specifically, as the YSR resonance shifts towards the SC gap edge, the tip-molecule interaction reduces the coupling of the magnetic moment with the superconducting substrate. Speculating, we are likely to be in the attractive force regime of the tip-sample interactions and as the molecule is weakly (van der Waals) bonded on the surface, the tip-sample interaction could have an effect on its adsorption height. Alternatively, the screening from the metallic tip could have an effect on the scalar potential at the impurity site, which would also have an effect on the YSR energy.

These experiments allow us to conclude that under our normal conditions (spectra in the manuscript are recorded in conditions similar to $z_\mathrm{offset}= 0$), the tip-sample interactions do not play a significant role when we carry out the d$I$/d$V$ spectroscopy in the middle of the molecule. However, as the tip moves towards the sample close to the edges of the molecule, some shifts of the YSR resonances may occur due to the tip-sample interaction.

\begin{figure}[!h]
	\centering
	\includegraphics [width=0.75\textwidth] {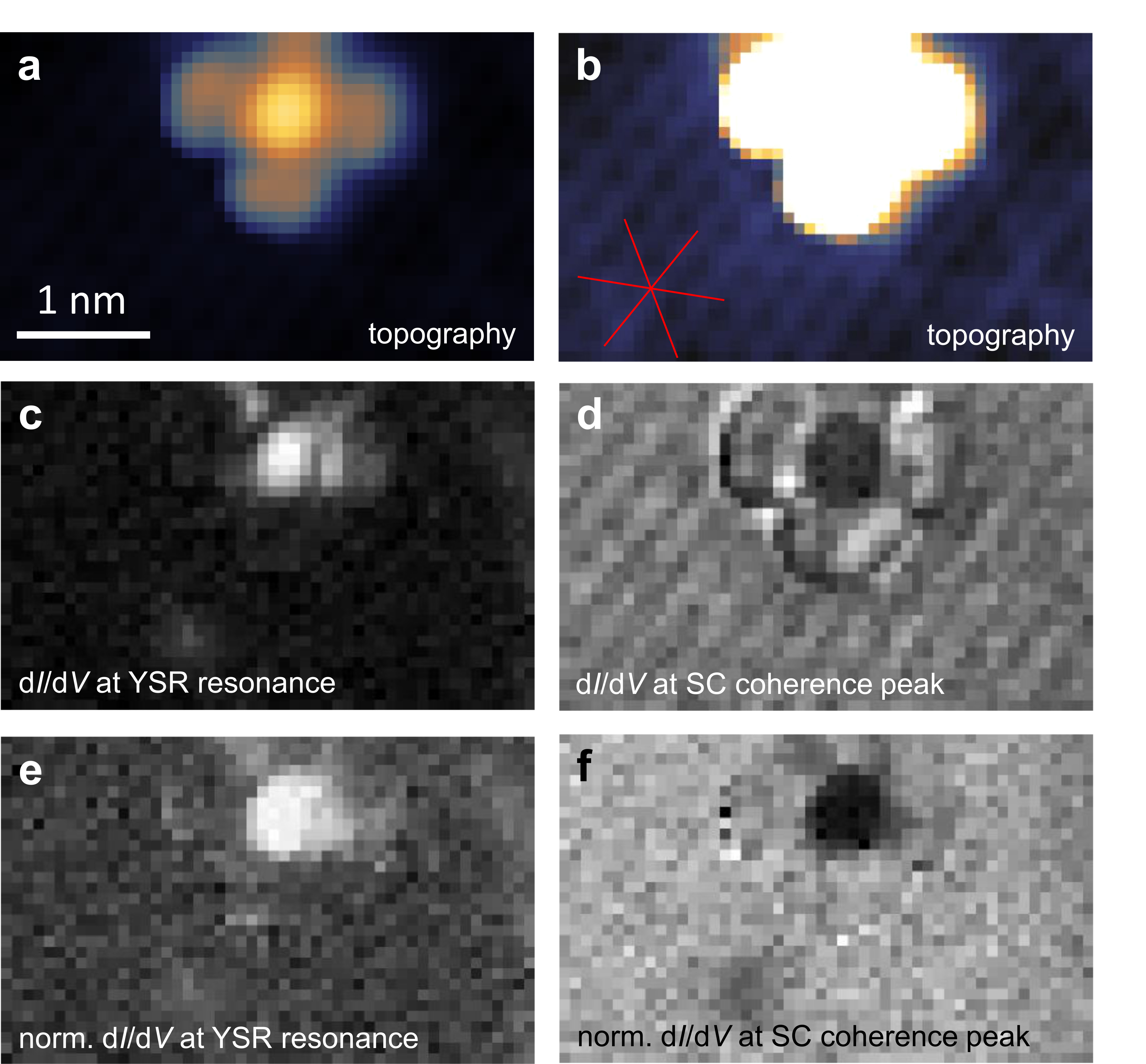}
	\caption{Grid spectroscopy on an individual CoPc molecule. (a) Topography image acquired simultaneously with the d$I$/d$V$ spectra. (b), Topography at enhanced contrast to show the atomically resolved underlying NbSe$_2$ lattice. (c) d$I$/d$V$ slice at the bias corresponding to the YSR peak (positive bias). Gray scale is between $0-0.03$ $\mu$S. (d) d$I$/d$V$ slice at the bias corresponding to the superconducting coherence peak (positive bias). Gray scale is between $0-0.02$ $\mu$S. (e) Maps corresponding to panels c and d, where the d$I$/d$V$ signal has been normalized by the current at the beginning of the spectrum. Set-point $V=500$ mV, $I=5$ pA, $z_\mathrm{offset}=100$ pm.}
	\label{Fig_grid}
\end{figure}

In order to shed further light into the YSR states on single CoPc molecules, we have mapped them out by performing grid spectroscopy (recording a complete d$I$/d$V$ spectrum at each scan point). These experiments are quite demanding due to the mobility of the molecules and we carried them out in STM feedback to have enough signal on the substrate and not to have too much current on the molecule. In addition, we used as low currents as possible to minimize tip-molecule interactions and to reduce the changes of accidental lateral manipulation of the molecule. The results are shown in Fig. \ref{Fig_grid}, which displays raw data d$I$/d$V$ slices at the energies corresponding to the YSR (panel c) and the superconducting coherence peaks (panel d) and the same results normalized to take into account the varying tip height (panels e and f). 

The YSR peaks are mostly localized in the centre of the molecule (Fig.~\ref{Fig_grid}e), with faint tails in different directions that seem to coincide with the principal lattice directions of the underlying NbSe$_2$ substrate and not with the molecular symmetry. There is a corresponding dip in the SC coherence peak intensity (Fig.~\ref{Fig_grid}f). While the normalization removes most of the effect of the molecular backbone (which will has an effect on the tunneling barrier between the tip and sample), this is still faintly visible in (Fig.~\ref{Fig_grid}e and f, which complicates quantitative analysis of the d$I$/d$V$ maps.

\section*{STM manipulation}
\begin{figure}[!h]
	\centering
	\includegraphics [width=0.45\textwidth] {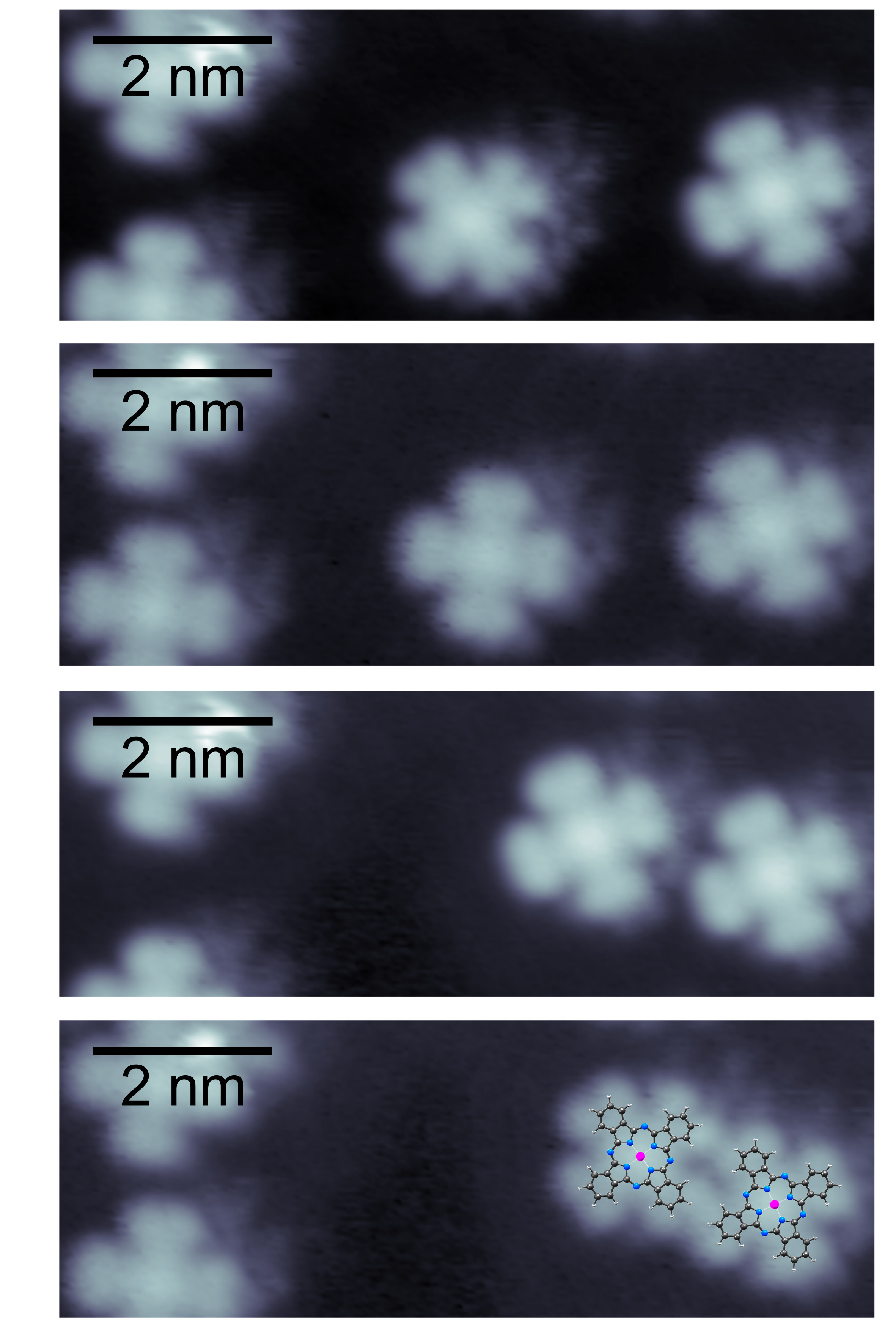}
	\caption{Lateral molecular manipulation of CoPc on NbSe$_2$. Subsequent STM images of CoPc molecules after one of the molecules was laterally manipulated by the STM tip.}
	\label{FigSI1}
\end{figure}
CoPc molecules are weakly adsorbed on the NbSe$_2$ surface, making it easy to laterally manipulate them by the STM tip (Fig. \ref{FigSI1}. We successfully constructed molecular dimers with different separations by STM manipulation. Manipulation of the CoPc was carried out by placing the tip above the centre of the molecule at 0.1 V bias voltage and increasing the current to 1 nA with the feedback engaged. The tip was then dragged towards the desired location. In order to avoid the potential variability caused by the adsorption orientation, we always  manipulated one molecule of the dimer and recorded the spectrum on the one which had not been moved. In this way, we have made sure that the target molecule is always in the same adsorption configuration. Figure \ref{FigSI1} shows a series of manipulation steps demonstrating that we only move the target molecule on the surface. The structural models of the CoPc molecules overlaid on the last panel show that the molecules are still not in "contact". 

\section*{Deconvolution of the \lowercase{d}$I$/\lowercase{d}$V$ spectra}
\begin{figure}[!h]
	\centering
	\includegraphics [width=0.4\textwidth] {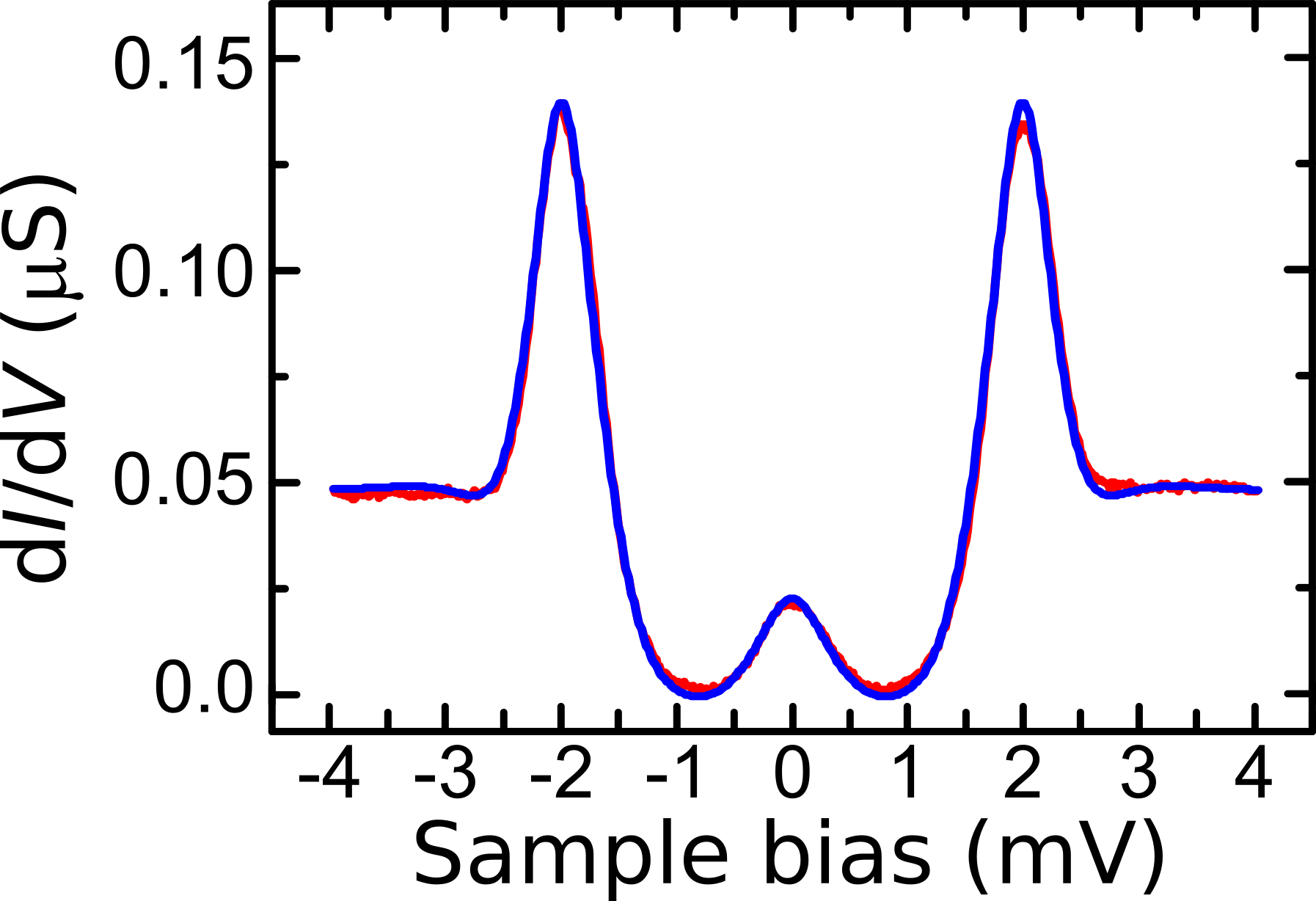}
	\caption{Deconvolution of the superconducting tip DOS. Experimental d$I$/d$V$ spectrum (red line) compared with the fit to Eq. \eqref{current} (blue line) with $\Delta_1=0.98$ meV, $\Delta_2=1.01$ meV,  $\gamma_1=0.16$ meV,  $\gamma_2=0.27$ meV,  $A_1=0.659$,  $A_2=0.341$, and $T_\mathrm{eff}=4.27$ K. Feedback set-point $V=100$  mV, $I=50$  pA, $z_\mathrm{offset}=100$ pm.}
	\label{fig:tip}
\end{figure}

The tunneling current $I$ at bias $V$ can be calculated from \cite{Tersoff1985}
\begin{equation}\label{current}
I(V)=\int_{-\infty}^{\infty} \rho_\mathrm{t}(\epsilon)\rho_\mathrm{s}(\epsilon+eV)\Big(f(\epsilon)-f(\epsilon+eV) \Big)d\epsilon
\end{equation}
where $\rho_\mathrm{t}$ and $\rho_\mathrm{s}$ are the tip and substrate densities of states and $f$ is the Fermi function. NbSe$_2$ has an anisotropic gap structure \cite{hudson-thesis}, which we approximated by sum of gapped DOS with some broadening and an additional gaussian component, similar to the expressions used before for modelling STM experiments with SC tips \cite{Franke2011,Ruby2015b}
\begin{equation}
\rho_\mathrm{SC}(E)=A_1\mathrm{Re}\left(\frac{|E|}{\Big((|E|+i\gamma_1)^2-\Delta_1^2\Big)^{1/2}}\right)+A_2\exp(-(|E|-\Delta_2)^2/(2\gamma_2^2))
\end{equation}
where $A_{1,2}$, $\Delta_{1,2}$ and $\gamma_{1,2}$ are fitted from the spectra measured on a clean NbSe$_2$ substrate with a superconducting tip. We can fit the experimental spectra extremely well as shown in Fig. \ref{fig:tip} using the same parameters for the bulk NbSe$_2$ substrate and for the superconducting tips prepared by controlled contacts with the clean substrate. After this, using the same tip with a known $\rho_\mathrm{t}$, we extract the substrate DOS on the CoPc molecules by a direct numerical deconvolution of Eq. \eqref{current}. This is an example of a Fredholm integral equation of the first kind and can be solved numerically through a matrix equation \cite{Twomey1963}. 

\section*{Theoretical description of coupled magnetic moments}

\subsection{Lattice model}
In this supplement, we outline in detail the theoretical modelling of coupled Yu-Shiba-Rusinov states on the surface of NbSe$_2$. The starting point of the analysis is a lattice description of an $s$-wave superconductor with Hamiltonian 

\begin{equation}
H_0=\sum_{ij,\sigma}\frac{t_{ij}}{2}(c_{i\sigma}^{\dagger}c_{j\sigma}+c_{j\sigma}^{\dagger}c_{i\sigma})+\sum_{i,\sigma}\Delta(c_{i\sigma}c_{i\bar{\sigma}}+c_{i\bar{\sigma}}^\dagger c_{i\sigma}^\dagger),
\end{equation}
where $t_{ii}$ is an on-site potential and $t_{ij}$ ($i\neq j$) are hopping elements between the lattice sites. The second term describes superconducting pairing of electrons with the pairing gap $\Delta$. The operators $c_{i\sigma}^{\dagger},c_{j\sigma}$ create and destroy electrons at lattice site $i$ with spin $\sigma$ and obey the usual fermionic anticommutation relations. We assume a triangular lattice where the band structure of NbSe$_2$ can be reproduced by an appropriate choice of parameters $t_{ij}$. The magnetic impurities are assumed to be local and described by the Hamiltonian 
\begin{equation}
H_{\mathrm{imp}}=-J_1{\bf S}_1\cdot C_{n_1}^{\dagger}\sigma C_{n_1}-J_2{\bf S}_2\cdot C_{n_2}^{\dagger}\sigma C_{n_2}+V_1 C_{n_1}^{\dagger}C_{n_1}+V_2C_{n_2}^{\dagger} C_{n_2},
\end{equation}
where $J_i, {\bf S}_i$ are a magnetic coupling and a classical spin vector of an impurity at lattice position $n_i$. Here, we have introduced Pauli matrices $\sigma=(\sigma_x,\sigma_y,\sigma_z)$ and the second-quantized spinor operators $C_{n_i}=(c_{i\uparrow},c_{i\downarrow})^T$. In addition to the magnetic coupling, the impurities may also perturb the superconductor with additional scalar potential parametrized by $V_1$ and $V_2$.

The standard method of solving the eigenstates of the Hamiltonian $H_0+H_{\mathrm{imp}}$ is to generalize the problem to particle-hole space with the basis $\Psi_{i}=(c_{i\uparrow},c_{i\downarrow},c_{i\downarrow}^\dagger,-c_{i\uparrow}^\dagger)^T$ and to diagonalize the Bogoliubov-de Gennes Hamiltonian  
\begin{eqnarray}
H_{\mathrm{BdG}}=\frac{t_{ij}}{2}\tau_z\otimes I_{2\times2}+\Delta\delta_{ij}\tau_x\otimes I_{2\times2}&-& J_1\delta_{n_1n_1}I_{2\times2}\otimes{\bf S}_1\cdot \sigma-J_2\delta_{n_2n_2}I_{2\times2}\otimes{\bf S}_2\cdot \sigma+ \nonumber \\
&V_1&\delta_{n_1n_1}\tau_z\otimes I_{2\times2}+V_2\delta_{n_2n_2}\tau_z\otimes I_{2\times2},
\end{eqnarray}
which is a $4N_1\times 4N_2$ matrix where $N_1$ and $N_2$ are the number of lattice sites in the direction of the primitive lattice vectors of a triangular lattice. We have solved the Bogoliubov-de Gennes problem $H_{\mathrm{BdG}}\Psi=E\Psi$ for a  tight-binding model of NbSe$_2$ with a finite on-site, nearest-neighbour and next-nearest neighbour hoppings. The energy spectrum is symmetric with respect to zero with a gap $2\Delta$. Inside the gap we recover two pairs of states, corresponding to the bonding and antibonding combinations of individual YSR states. In the calculations presented in the main text, we have used values of -100 meV for the on-site and -125 meV for the nearest- and next-nearest neighbour hopping parameters \cite{Flatt2000} and $\Delta=1$ meV, $J_1S_1= J_2S_2=35$ meV, $V_1=V_2=0$ on a lattice with $N_1=N_2=400$.  

\subsection{Continuum model}
\begin{figure}[!h]
	\centering
	\includegraphics [width=0.8\textwidth] {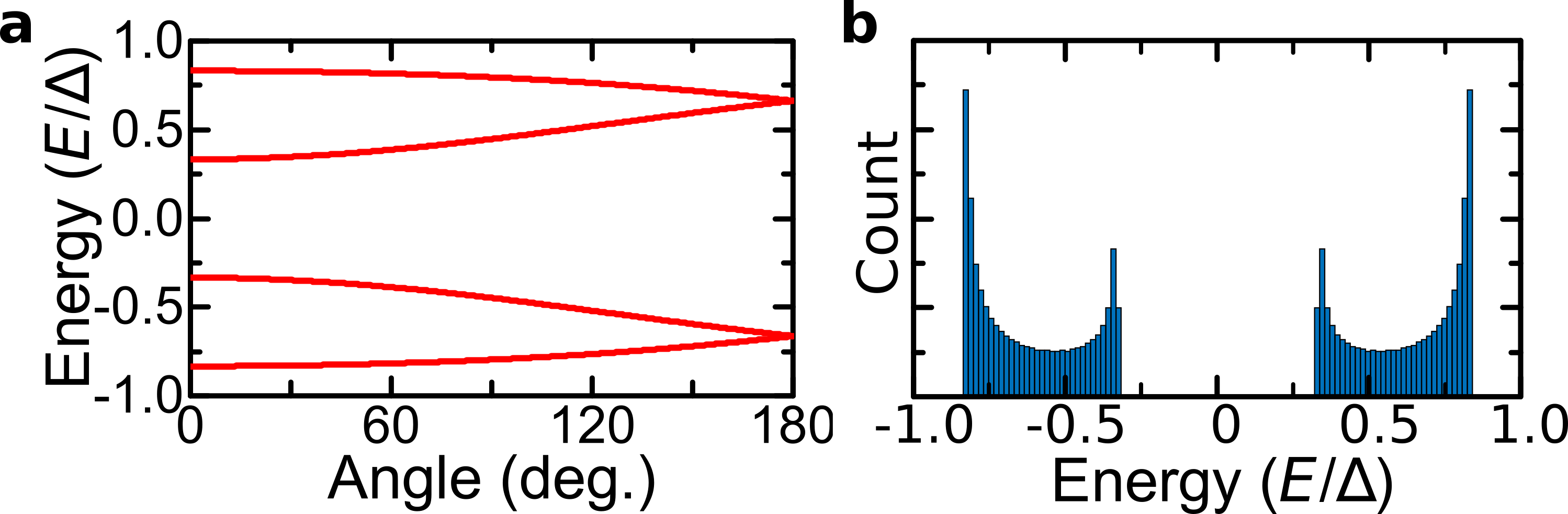}
	\caption{Angular averaging of the YSR splitting. (a) YSR state energies as a function of the angle of the magnetic moments of the two impurities according to the continuum model ($k_\mathrm{F}a = 3.8$, $\alpha=0.5$). (b) Same data presented as a histogram, which shows that majority of the angles result in a YSR splitting very close to the maximum value. This would result in clearly split resonances in the experimental spectra.}
	\label{FigSI3}
\end{figure}

To gain qualitative insight of the bound states of the coupled magnetic impurities we have also studied a 2D continuum model with a circular Fermi surface. In the continuum description the Bogoliubov-de Gennes Hamiltonian takes the form 
\begin{equation}
H_{\mathrm{BdG}}=\varepsilon_k\tau_z\otimes I_{2\times2}+\Delta\tau_x\otimes I_{2\times2}-J\delta({\bf r-r}_1)I_{2\times2}\otimes{\bf S}_1\cdot \sigma-J\delta({\bf r-r}_2)I_{2\times2}\otimes{\bf S}_2\cdot \sigma,
\end{equation}
where $\varepsilon_k=\frac{k^2}{ 2m}-\mu$ is the kinetic energy measured from the Fermi surface while the other terms are straightforward counterparts of those present in the lattice model. In the case of a single impurity, the standard calculation leads to a pair of subgap states with energies $E=\frac{1-\alpha^2}{1+\alpha^2}$, with the dimensionless coupling $\alpha=\pi\nu JS$ which contains the density of states at the Fermi level $\nu$. Since we are not including a spin-orbit coupling, the single impurity results does not depend on the orientation of the impurity moment. While the two-spin problem does not admit a simple closed form analytical expression, the eigenstates can be solved, for example, by the methods of Ref. \cite{Lutchyn2015}. This leads to four subgap states $\pm E_1$ and $\pm E_2$ where the energy splitting $|E_1-E_2|$ depends on the relative angle of the two impurity moments. The splitting is maximal for parallel moments and vanishes for antiparallel moments. As illustrated in Fig.~\ref{FigSI3}, the intermediate cases interpolate between the two cases in a nonlinear manner. Thus one expects a four-peak structure even when the spins are randomly oriented.  Due to the slow decay $\sim \frac{1}{\sqrt{r}}$ of 2D Shiba wavefunctions for distances smaller than the superconducting coherence length, the splitting decays slowly. Finally, as the experiments and the lattice calculations reveal, the isotropic Fermi surface cannot adequately capture coupling of two moments which varies strongly in a lattice scale. 

\subsection*{Angular averaging of the YSR splitting}

\end{document}